\documentclass[aps,prb,preprint,superscriptaddress,reprint,preprintnumbers,amssymb]{revtex4-1}

\usepackage{chemist}
\usepackage{xcolor}
\usepackage{graphicx}
\usepackage{units}
\usepackage[version=3]{mhchem}

\newcommand{\paragraphlb}[1]{\paragraph{#1}\mbox{}\\}

\newcommand{\LG}[1]{#1}%\textcolor{red}{#1}}
\newcommand{\SM}[1]{#1}%\textcolor{red}{#1}}
\newcommand{\LR}[1]{#1}%\textcolor{red}{#1}}
\newcommand{\TD}[1]{#1}%\textcolor{red}{#1}}
\newcommand{\delete}[1]{}

\begin{document}

\title{Challenges in Large Scale Quantum Mechanical Calculations}

\author{Laura E. Ratcliff}
\affiliation{Argonne Leadership Computing Facility, Argonne National Laboratory, Illinois 60439, USA}
\author{Stephan Mohr}
\affiliation{Department of Computer Applications in Science and Engineering, Barcelona Supercomputing Center (BSC-CNS), Barcelona, Spain}
\author{Georg Huhs}
\affiliation{Department of Computer Applications in Science and Engineering, Barcelona Supercomputing Center (BSC-CNS), Barcelona, Spain}
\author{Thierry Deutsch}
\affiliation{Univ.\ Grenoble Alpes, INAC-MEM, L\_Sim, F-38000 Grenoble, France}
\affiliation{CEA, INAC-MEM, L\_Sim, F-38000 Grenoble, France}
\author{Michel Masella}
\affiliation{Laboratoire de Biologie Structurale et Radiologie, Service de Bio\'energ\'etique, Biologie Structurale et M\'ecanisme,Institut de Biologie et de Technologie de Saclay, CEA Saclay, F-91191 Gif-sur-Yvette Cedex, France}
\author{Luigi Genovese}
\email{luigi.genovese@cea.fr}
\affiliation{Univ.\ Grenoble Alpes, INAC-MEM, L\_Sim, F-38000 Grenoble, France}
\affiliation{CEA, INAC-MEM, L\_Sim, F-38000 Grenoble, France}

\date{\today}
\begin{abstract}
During the past decades, quantum mechanical methods have undergone an amazing transition from pioneering investigations of experts into a wide range of practical applications, made by a vast community of researchers. First principles calculations of systems containing up to a few hundred atoms have become a standard in many branches of science. 
The sizes of the systems which can be simulated have increased even further during recent years, and quantum-mechanical calculations of systems up to many thousands of atoms are nowadays possible.
This opens up new appealing possibilities, in particular for interdisciplinary work, bridging together communities of different needs and sensibilities.
In this review we will present the current status of this topic,
and will also give an outlook on the vast multitude of applications, challenges and opportunities stimulated by electronic structure calculations, making this field an important working tool and bringing together researchers of many different domains.
\end{abstract}
\maketitle
\section{Introduction}
The fundamental laws for a Quantum Mechanical (QM) description of atomistic systems up to 
the nanoscale are known and have been well established for 
a little less than a century.
Yet, there are many challenges related to the Quantum Mechanical treatment of large systems.
In the vast majority of cases, we are still unable to solve the fundamental Schr\"odinger equation for systems of realistic sizes in such a way that the results satisfy ``universal'' requirements of accuracy, precision and especially predictability.
Unfortunately, this also implies that we are still far from being able to quantitatively predict  experimental results at the nanoscale.
%Quantum Mechanics (QM) methods are challenging in Quantum Chemistry (QC) since the second half of last century.

The problems are not only related to the computational complexity needed to solve the equations of QM, there are also intrinsic obstacles.
To give an example, let us remind the so-called ``Coulson's challenge''.
In 1960, Coulson~\cite{Coulson1960a} noticed that the most compact object needed to 
characterize quantum mechanically an $N$-electron system (at least in its ground state) is the two-body reduced density matrix (2RDM).
However, it turns out that \emph{we do not know} all the necessary conditions for the 2RDM to be \emph{$N$-representable}, 
i.e.\ coming from an anti-symmetric wavefunction of an $N$-electron system.
Thus, even if a compact (and, in principle, computationally accessible) object exists, theoretical and algorithmic bottlenecks hinder its practical usage.
In 1964~\cite{Hohenberg-Kohn.1964} and 1965~\cite{Kohn-Sham.1965}, Kohn, Hohenberg and Sham further reduced the complexity by showing that
the electronic density is in a one-to-one correspondence with the ground state energy of a system of interacting electrons,
and that such an interacting system can be replaced by a mean-field problem of $N$ non-interacting 
fermions that provide the same distribution of the density. These are the fundamental ideas of Density Functional Theory (DFT).

\LG{DFT has been, for more than twenty years,
the workhorse method for simulations within the solid state community. 
Moreover, in spite of the fact that DFT drastically reduces the complexity with respect to the \emph{ab initio} methods of Quantum Chemistry,
the success of such a treatment in the latter community is undeniable. This is mainly due to the fact that,
on one hand, the quality of the exchange and correlation functionals available permits the calculation of certain properties
with almost chemical accuracy, and, on the other hand, there are numerous software packages, which are relatively easy to use,
that have contributed to the diffusion of the computational approach.}

\delete{In spite of this enormous reduction in the complexity, DFT maintains 
a fair level of first principles description and accuracy. Thus, it has been, for more than twenty years,
the workhorse method for simulations within the solid state community, and it is nowadays widely accepted in 
Quantum Chemistry as a reasonable compromise between the quality of the description and the computational cost.
During the past years the typical sizes of the systems that can been treated has considerably increased. 
Since more or less one decade ago, we have entered the ``second era'' of DFT, 
where calculations of higher complexity and of systems of larger sizes are possible, in this way also extending the range of possible applications to new fields.}

\LG{During the past years, there has been a multiplication of DFT software packages that are able to
treat systems of increasingly large size.}
This has, on one hand, been enabled by both the advances in supercomputing architectures and the code developers continuously improving their codes to exploit the steadily increasing performance provided, but it is at the same time also motivated by various scientific needs.
\LG{This fact clearly extends the range of possible applications to new fields, and to communities
traditionally focused on larger systems.
In a similar manner to the uptake of DFT in the Quantum Chemistry community, things are progressing as if we are entering a ``second era'' of DFT calculations,
where DFT and, more generally, large-scale quantum mechanical treatments, are susceptible to wide diffusion in other communities.}

In this review paper we will present some of the motivations that led computational physicists and quantum chemists into this second era. 
The different aspects will be separated into various subcategories, while trying to give a general overview, and will be completed by notable examples in the literature.
This inspection of the state-of-the-art will provide the reader with an outlook on the present capabilities of QM approaches.
We will then continue our discussion by presenting the key concepts that have emerged in the last decade.
These concepts are often specific to QM calculations at large scale and are rather different from those which are typical of traditional calculations, where the systems' sizes are \LG{limited to a few hundreds of orbitals}. \delete{only a few tens of atoms.}
These concepts are therefore of high importance for potential users of such advanced DFT methods. 

\subsection{The need for large-scale QM}
Given the unbiased predictive power of Quantum Mechanics, there is obviously no need to explain why systems containing only a few atoms should be modelled using this approach.
However, for simulations of systems at the nanoscale, composed of many thousands of atoms, the question of the need for a Quantum Mechanical treatment might appear legitimate. For systems of these sizes, the electronic degrees of freedom are seldom of interest, and the interatomic potential might be described by more compact approaches like Force Fields, possibly fine-tuned for describing experimentally known structural and dynamical or polarizability properties.
In other terms, the intimate nature of the problem changes: instead of focussing solely on the correct estimation of interactions and correlation between electrons, one rather has to concentrate on the exhaustive \emph{sampling} of the configuration space, thereby losing the need for an intrinsic quantum mechanical description.

In addition we know that, even if a QM approach were feasible, this would not necessarily lead to a better description. Although the complexity of the model is certainly higher and the description is less biased, there are still many approximations which are hidden in a Quantum Mechanical calculation; therefore we are---even for a system containing only a handful of atoms---in general still far from chemical accuracy. The situation for large systems will be the same or even worse, since more severe approximations have to be adopted.
%does not necessarily provide results closer to the experiment: lots of approximations are hidden, and we are still far from chemical accuracy for few-atom systems, therefore the situation for large system would not improve.
However, the need for QM calculations of large systems does not solely come from a quest for accuracy. Indeed there are other reasons why an \emph{ab initio} description for large systems is desirable or even crucial, and one of the purposes of this review paper is to identify and discuss some of these aspects.

\begin{figure*}
\includegraphics[width=0.75\textwidth]{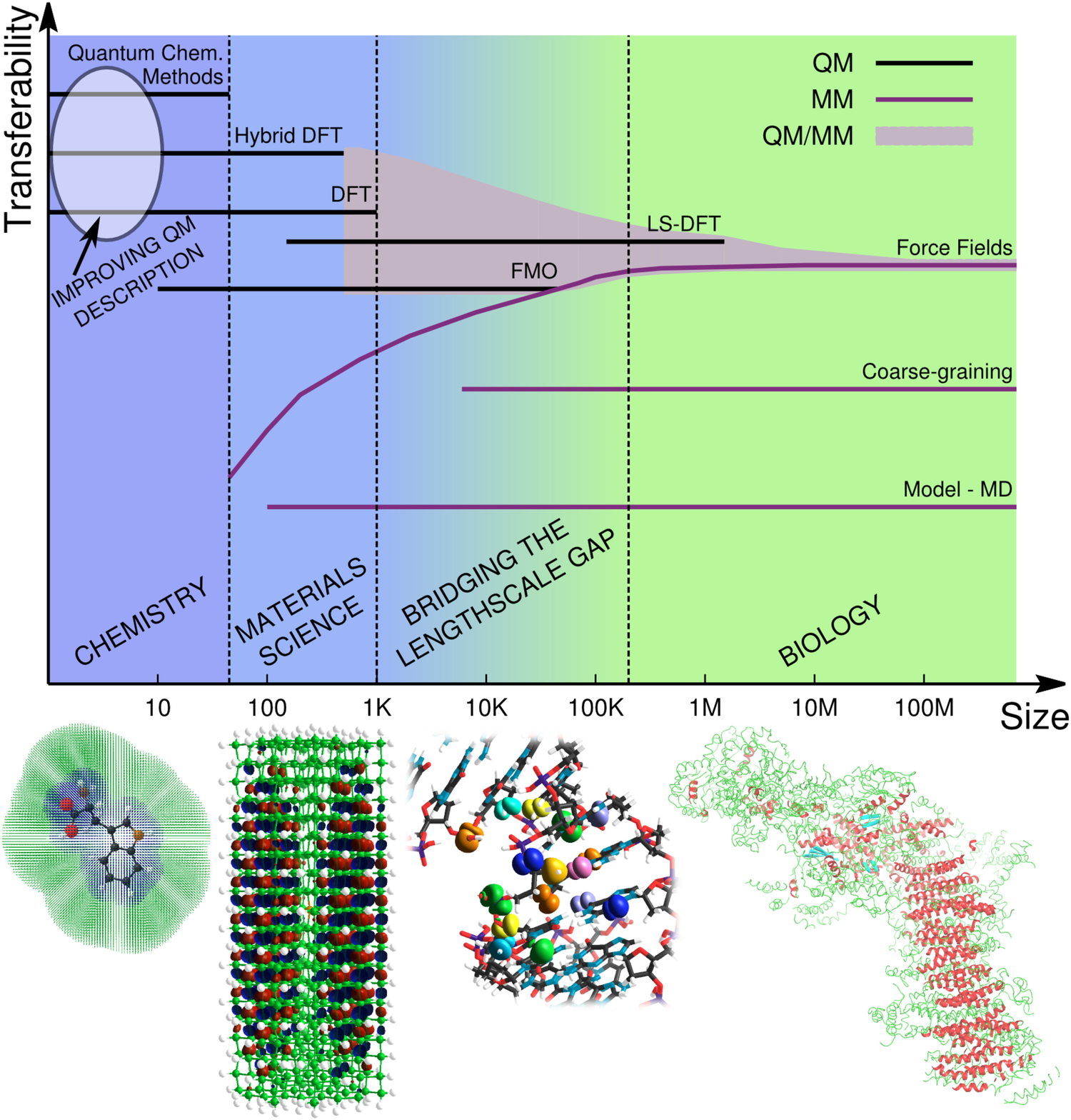}
\caption{\delete{Overview of the typical length scales and precisions for popular methods used in simulations of systems atomistic systems at the mesoscale.}
\LR{Overview of the popular methods used in simulations of systems \LG{with atomistic resolution}, showing the typical length scales over which they are applied as well as the degree of transferability of each method, i.e.\ the extent to which they give accurate results across different systems without re-tuning.}
 On the left hand side we have the \delete{highly accurate} Quantum Chemistry methods which are \LR{highly transferable but only applicable to a few tens of atoms}\delete{ mainly applicable to some tens of atoms}; on the right hand side we see the less \delete{accurate} \LR{transferable} (semi-)empirical methods, which can however express reliable results (as they are parametrized for) for systems containing millions of atoms; and in the middle we see the methods---in particular linear-scaling DFT---which can bridge the gap between the two regimes.  
\LR{The vertical divisions and corresponding background colors give an indication of the fields in which the methods are typically applied, namely chemistry, materials science, biology and an intermediary regime (`bridging the length scale gap') between materials science and biology.  The line colours indicate whether a method is QM or MM, while the typical regime for QM/MM methods is indicated by the shaded region.
In the top left the region wherein efforts to improve the quantum mechanical treatment are focussed, that is the quest to climb `Jacob's ladder' by developing new and improved exchange correlation functionals, is also highlighted.}
Some representative systems for the different regimes are depicted along the bottom: the amino acid tryptophan with a multi-resolution grid, a defective Si nanotube with an extended KS wavefunction, DNA with localized orbitals, and the protein mitochondrial NADH:ubiquinone oxidoreductase~\cite{Zickermann2015}.}
%\textcolor{red}{[SM: This is the accession code in the Protein Data Bank... Is this ok?]}.}
\label{QMill}
\end{figure*}

The present-day scenario of the available methodological techniques to study systems at the atomistic level
can be sketched in Fig.~\ref{QMill}: here various methods are illustrated within the typical scales where they
have been usually applied.
It is interesting to notice that the size where typical QM approaches are developed and improved is of the order of few atoms, even though some of these concepts are then also applied to larger systems.
\LG{John Perdew introduced the renowned metaphor of \emph{Jacob's Ladder}~\cite{PerdewLadder},}
\delete{For systems of interest in Quantum Chemistry, a renowned illustration is provided by the so-called \emph{Jacob's Ladder}~\cite{PerdewLadder}
of DFT approximations,}
where the computational complexity of the implementation of the \LG{DFT} exchange and correlation functional is (in principle) 
directly related to the accuracy of the description, aiming at the ``heaven'' of chemical accuracy.

Likewise, for more than ten years, a lot of work has been done to extend the range of applicability of QM methods to larger systems.
\LG{The so-called \emph{nearsightedness property}\cite{KohnNS} suggested that, at least in principle, 
one could exploit locality to build linear scaling methods that are able to reach larger scales.}
Initially\LG{, the development of such computational methods was driven by the ``academic'' purpose of verifying the computational consequences of 
nearsightedness.} \delete{ this investigation has been driven by academic purposes, following the so-called }
In Sec.~\ref{seclarge} we will overview the most important advancements in this topic and some of the established computational
approaches in large-scale QM. This is by no means new and there are a number of valid review papers on the topic, to which we will also refer. Our aim is not to be fully exhaustive on this topic as there has been many research studies in this direction.
However we would like to put the emphasis on the fact that nowadays the panorama is so rich and there is enough diversity in the 
computational approaches to claim that such a discipline is now mature enough to be largely diffused also among non-specialists.

The reason for this diffusion is related to the \emph{opportunities} that a QM approach opens for systems composed of thousands, if not hundreds of thousands, atoms.
On the one hand, there are quantities which are intrinsically only accessible using QM, for example all investigations dealing with electronic excitations~\cite{Lille1};
%we will presen For instance, all investigations dealing with electronic excitations require a QM treatment; 
%\textcolor{blue}{[GH: double formulation, my suggestion: "...only accessible using QM, for example all investigations dealing with electronic excitations; we will present...". SM: much better, corrected]}
we will present some more examples in Sec.~\ref{secQMMM} and~\ref{extmossec}.
On the other hand, large QM calculation are also needed to access \emph{error bars and statistics} of the results. An example is the need to get good statistics
among different constituents in a morphology---a task which is not possible by implicit, 
classical modelling of the environment.
In addition, another aspect where first-principles QM approaches are important is the need for \emph{validation} of non-QM approaches.

For all of these tasks, there is a typical length scale, ranging from a few hundred to many thousands of atoms, where it is
important to master \emph{both} QM and classical approaches.
\LG{As discussed in the introduction, it was not possible in the initial implementations of DFT software packages---for various reasons, including the available computational resources--- 
to reach such large length scales, i.e.\ there was a ``length scale gap'' between the maximum scale which was accessible to QM and the typical scale at which classical approaches are applied.}\delete{In the early days of QM methods and DFT in particular, there has been a ``length scale gap''}
\LG{QM computational paradigms
had to bridge this gap in order to be used as investigation tools for systems of many thousand atoms.}\delete{between QM computational paradigms 
and classical approaches
where it was not possible to use both paradigms at the same time.}
However, since a few years ago and mostly driven by the development of linear-scaling QM methods, this gap has vanished.
%there is not anymore a ``lenghtscale gap'' between the two investigation paradigms.
Thus, intensive research and investigation in this range will allow the set up of new, powerful computational approaches in various disciplines such as soft matter, biology and life sciences.

%We believe that it is timely to highlight them, 
%as such a discipline is now mature enough to be largely diffused also among non-specialists.

\section{Large scale QM: Methodological and computational approaches}\label{seclarge}

The problem of treating large systems with DFT is not a new one; indeed research into this area goes back more than two decades~\cite{Galli1996,Goedecker1999}.  This work focused on developing new methods with reduced scaling, leading to the different linear-scaling DFT (LS-DFT) codes which exist today. The emphasis was initially on academic interest, that is to say the focus was on the methods themselves and finding new and better ways to accelerate calculations of ever larger systems, rather than on the application to \LG{major} scientific problems.  Indeed, until more recently, the vast majority of applications were limited to proof-of-concept calculations, which served to demonstrate the capability of these algorithms to treat ever larger systems, while hinting at future possibilities for production calculations.  Nonetheless, without this pioneering work, we would not be in the position today to tackle large and challenging systems such as those discussed in more detail below.

The development of reduced scaling methods was also naturally coupled with the availability of high performance computing resources; thanks to both the increase in computing power of the fastest supercomputers and the widespread availability of commodity clusters, LS-DFT can now not only be applied to very large systems indeed, but it can also do so while maintaining the same accuracy as more traditional cubic-scaling approaches.    

As a result of the complexities involved in such methods, their usage was initially mostly limited to experts within the community.  This is no longer entirely the case, however there remain a number of additional concepts with which interested users must familiarize themselves before attempting practical calculations.   In this section we give an overview of some of these important concepts, notably the quantum-mechanical principle of nearsightedness, which provides the justification for linear-scaling methods, and the codes within which they are implemented.  This is not intended to be a fully exhaustive list, rather the aim is to highlight the most popular approaches and some of the key achievements within the field.  For a more thorough discussion, the reader is encouraged to refer to other, more extensive reviews of the subject~\cite{Galli1996,Goedecker1999,Bowler2002,Bowler2012}.

\subsection{Nearsightedness and linear scaling}
\delete{In principle quantum mechanics is a non-local concept, as can be seen for 
instance by the antisymmetry condition which a many-electron wave function must 
fulfil, no matter how far away the electrons are from each other.} 
In the context of DFT, \LG{the tendency of the Kinetic Energy operator to favour the 
delocalisation of the Kohn-Sham orbitals means that they} 
\delete{,the most popular \emph{ab initio} method for solving an electronic 
structure problem, this translates to the extent of the Kohn-Sham (KS) orbitals, 
which} are in general extended over the entire system. This non-locality leads to 
an unfavorable cubic scaling, meaning that an increase of the system size by a 
factor of ten leads to a
computational effort which is 1000 times greater. Even though this is 
considerably better than the scaling of other popular Quantum Chemistry methods, 
which ranges from $\mathcal{O}(N^4)$ for Hartree Fock (HF) to $\mathcal{O}(N^5)$ 
for MP2, $\mathcal{O}(N^6)$ for MP3 and $\mathcal{O}(N^7)$ for MP4, CISD(T) and 
CCSD(T), it still makes large scale simulations prohibitive.

On the other hand, the density matrix $F(\mathbf{r};\mathbf{r}')$, which is an integrated quantity that is invariant under unitary transformations of the Kohn-Sham orbitals, does not reflect this non-locality. Indeed it can be shown that the elements of the density matrix decay rapidly with respect to the distance between $\mathbf{r}$ and $\mathbf{r}'$: for insulators and metals at finite temperature exponentially~\cite{cloizeaux1964energy,cloizeaux1964analytical,kohn1959analytic,baer1997sparsity,ismail-beigi1999locality,goedecker1998decay,he2001exponential}, and for metals at zero temperature algebraically~\cite{march1967the}. Kohn has coined the term ``nearsightedness'' for this effect~\cite{kohn1996density}, and this concept is the key towards calculations of very large systems: by truncating elements beyond a given cutoff radius it is possible to reach an algorithm which scales only linearly with respect to the size of the system. 

\begin{figure}
 \includegraphics[width=0.21\textwidth]{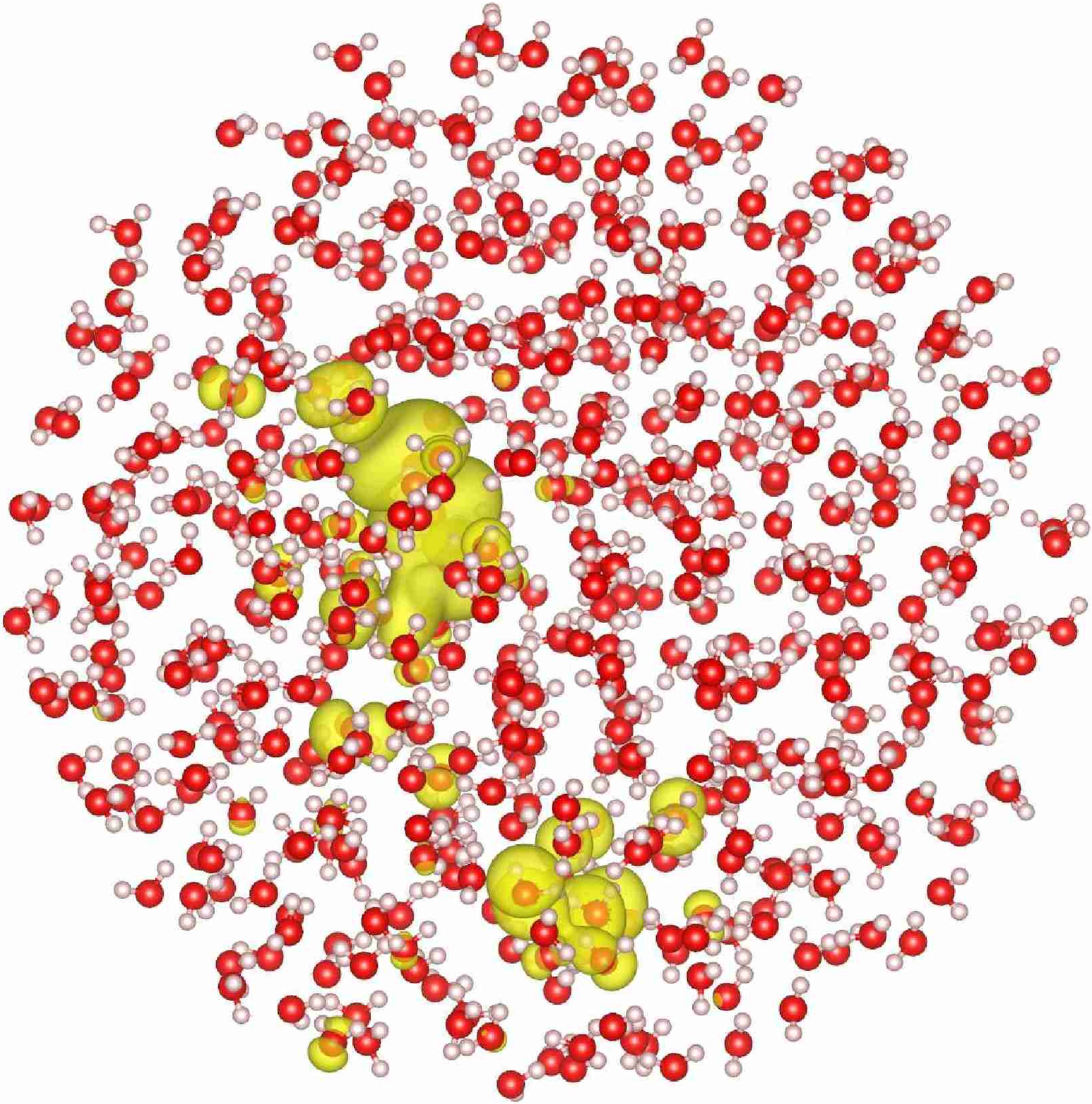}
 \includegraphics[width=0.25\textwidth]{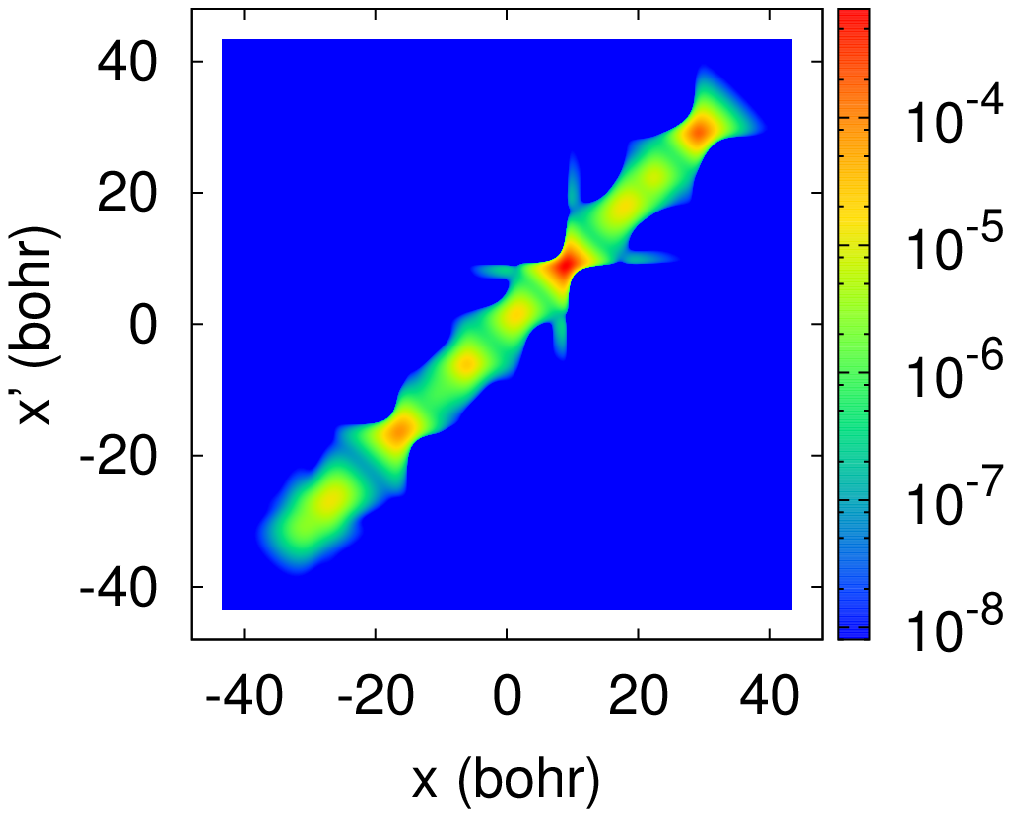}
 \caption{Left: Isosurface of one Kohn-Sham orbital for a water droplet consisting of 1500 atoms. Right: density matrix in the $x$ dimension, i.e.\ $F(x,y_0,z_0;x',y_0,z_0)$, for the same system.}
 \label{fig:water_decay_properties}
\end{figure}

An illustration of this effect is shown in Fig.~\ref{fig:water_decay_properties} for the case of a water droplet containing 1500 atoms. Here we plot on the left side the isosurface of an extended Kohn-Sham orbital, and on the right side the density matrix of the system in the $x$ dimension, i.e.\ $F(x,y_0,z_0;x',y_0,z_0) = \sum_i f(\epsilon_i) \psi_i(x,y_0,z_0)\psi_i(x',y_0,z_0)$, where $\psi_i$ are the Kohn-Sham orbitals and $f(\epsilon_i)$ their occupation numbers. As can be seen, the summation of the extended orbitals nevertheless leads to a localized quantity, meaning that the non-local contributions are cancelled due to interference effects.

In a linear-scaling DFT approach, this locality must be taken advantage of, which can be achieved by building the algorithm directly on the density matrix, rather than the Kohn-Sham orbitals.
Since this may introduce an additional computational overhead, these $\mathcal{O}(N)$ algorithms are usually slower for small systems than traditional approaches and only outperform the latter ones beyond a critical system size, the so-called crossover point.  This crossover point is dependent not only on the details of the method used, but also the properties, in particular the dimensionality of the system being studied.  In many linear-scaling DFT approaches, such as \textsc{onetep}~\cite{Skylaris2005}, \textsc{Conquest}~\cite{Bowler2010}, \textsc{Quickstep}~\cite{VandeVondele2005} and \textsc{BigDFT}~\cite{mohr-daubechies-2014,mohr-accurate-2015}, the density matrix is written in separable form as
\begin{equation}
 F(\mathbf{r},\mathbf{r}') = \sum_{\alpha,\beta} \phi_{\alpha}(\mathbf{r}) K^{\alpha\beta} \phi_{\beta}(\mathbf{r}') ,
\end{equation}
with a set of so-called support functions $\phi_{\alpha}(\mathbf{r})$ and the density kernel $\mathbf{K}$. In order to reach a linear complexity, the support functions are strictly localized and the density kernel is enforced to be sparse, meaning that elements are set to zero beyond a given cutoff radius.  Different approaches can be used to find the ground state density matrix, which are discussed below.

\subsection{Reduced-scaling approaches and established codes}

In the following we describe both pioneering early approaches to LS-DFT and modern, state of the art methods currently being used for applications.  Since this review is intended to be of practical use rather than purely theoretical, where appropriate, we categorize the various approaches by the code in which they are implemented.
It should be noted that many, though not all, of the approaches to LS-DFT described below are valid only for systems with a band gap, since, as mentioned above, the density matrix decays only algebraically at zero temperature for metals, rather than exponentially.  Exponential decay, is, however, recovered for metals at finite temperature, thereby providing one avenue for LS-DFT with metals.  Where relevant, we mention if the codes are capable of treating metallic systems.

\paragraphlb{Pioneering Order N methods}
The earliest LS-DFT method was the divide and conquer approach of Yang~\cite{Yang1991}.  As the name implies, in this approach the system of interest is divided into a number of smaller subsystems which can be treated independently using a local approximation to the Hamiltonian.  The KS energy for the full system is extracted from the subsystems, which are coupled via the local potential and Fermi energy.  Since the size of the subsystems is independent of the total system size, the method scales linearly, and is also straightforward to parallelize, however, the crossover point can be rather high.  

Another pioneering work proposed an order N method to calculate the density of states and the band structure by means of the Green function and a recursion method to calculate the moments of the electronic density~\cite{Baroni-Giannozzi.1992} in real space using a finite difference scheme. The evaluation of the moments of the electronic density by means of random vectors was also used to treat systems up to $2160$ atoms~\cite{Drabold1993}.
Many density matrix minimization methods were also developed; the Fermi Operator Expansion (FOE)~\cite{Goedecker1994,Goedecker1995}, LNV (Li-Nunes Vanderbilt)\cite{li5,lnv2} and other approaches related to the purification transformation~\cite{McWeeny1960} are currently used in various codes today.
The authors refer to the comprehensive review of S.~Goedecker~\cite{Goedecker1999} for a full description of these different methods.

\paragraphlb{SIESTA}
The first widely used code with a linear-scaling method was \textsc{SIESTA}~\cite{Artacho1999,siesta_on,siesta_website}, which is based on numerical atomic orbitals.  The original linear-scaling approach is based on the minimization of the functional of Kim, Mauri and Galli~\cite{Kim1995}, which avoids explicit orthogonalization.
More recently a divide-and-conquer algorithm has also been implemented~\cite{Cankurtaran2008}. There are many real applications using \textsc{SIESTA} for large systems, but these calculations use a traditional cubic-scaling scheme based on the diagonalization of the Hamiltonian. In 2000, $\lambda$-DNA of $715$ atoms was calculated using the linear-scaling method to show the absence of DC-conduction~\cite{DePablo2000}. In 2006, the calculation of some CDK2 inhibitors was done using \textsc{SIESTA}, with also a comparison to \textsc{onetep}~\cite{Heady2006}.
Recently \textsc{SIESTA} has been coupled~\cite{Lin2014} with the \textsc{PEXSI} library~\cite{Lin2013}, which avoids the cubic-scaling diagonalization of the Hamiltonian by taking advantage of its sparsity in the localized basis. This reduces the computational complexity without the need for nearsightedness or other simplifications, thus allowing considerably larger systems to be tackled without requiring any explicit truncation of the density matrix.
The first published scientific application of \textsc{SIESTA}-\textsc{PEXSI} examines carbon nanoflakes up to a size of 11,700 atoms~\cite{Hu2014}.
%LR: this sounds like it's still cubic-scaling, but it should be quadratic at most while remaining within the traditional cubic scaling scheme, i.e.\ not imposing any truncation.
In addition \textsc{SIESTA} allows one to perform electron transport calculations using the \textsc{TranSIESTA} tool~\cite{Stokbro2003} providing a tight binding Hamiltonian
and can also be used for QM/MM simulations~\cite{Sanz-Navarro2011}.

\paragraphlb{ONETEP} 
The Order-N Electronic Total Energy Package \textsc{onetep}\cite{Skylaris2005,Skylaris2008,Hine2009,onetep_website} is a LS-DFT code which employs a density matrix approach, wherein the strictly localized support functions, termed Non-orthogonal Generalized Wannier Functions (NGWFs), are represented in a basis of periodic sinc (psinc) functions and optimized \emph{in situ}, adapting themselves to the chemical environment.  Since the psinc basis can be directly related to plane-waves, the NGWFs form a localized minimal basis with the same accuracy as a plane-wave calculation.   The density kernel is calculated primarily using the LNV approach in combination with other methods~\cite{Haynes2008}. A number of functionalities have been implemented in \textsc{onetep} such as DFT+U~\cite{ORegan2012}, the calculation of optical spectra~\cite{Ratcliff2011}, including via time-dependent (TD) DFT~\cite{Zuehlsdorff2013,Zuehlsdorff2015}, constrained DFT~\cite{Turban2016}, \LR{electronic transport~\cite{bell-2015-electronic}}, natural bond orbital analysis~\cite{Lee2013}, and implicit solvents~\cite{Dziedzic2011}.  A method to treat metallic systems at finite temperature has also been implemented~\cite{Ruiz-Serrano2013}.

Many large calculations have been performed with \textsc{onetep}, such as on 
DNA (2606 atoms)~\cite{Skylaris2005}, 
carbon nanotubes (4000 atoms)~\cite{Hine2010}, 
a silicon crystal (4096 atoms)~\cite{Hine2011},
and point defects in Al$_2$O$_3$~\cite{Hine2010}.
\textsc{onetep} was also used in biology to study the binding process within a 1000-atom QM model of the myoglobin metalloprotein~\cite{Weber2014} and also, in a QM/MM approach, the transition state optimization of some enzyme-catalyzed reactions~\cite{Lever2014}. 
Some of the applications with \textsc{onetep} have clearly highlighted the need for and challenges associated with large scale QM calculations. For example there is a clear need for methods capable not only of incorporating and analysing electronic effects on a large scale in proteins including the solvent effect, at least implicitly, as demonstrated by the study of a 2615-atom protein-ligand complex~\cite{Dziedzic2011}, but also of optimizing transition state (TS) structures in this context. 
In another study, \textsc{onetep} was used to put in evidence the importance of preparing systems correctly to avoid the problem of the vanishing gap for large systems (proteins and water clusters)~\cite{Lever2013}.

%OMM!
  
\paragraphlb{OPENMX} %http://www.openmx-square.org  
The \textsc{OpenMX} code~\cite{Ozaki2006,openmx_website} has both a linear-scaling version, based on the divide and conquer approach defined in a Krylov subspace~\cite{Ozaki2006}, and a cubic-scaling version which uses diagonalization.  It uses a basis set of pseudo-atomic orbitals (PAOs) and a number of functionalities have been implemented, such as DFT+U~\cite{PhysRevB.73.045110}, electronic transport~\cite{PhysRevB.81.035116} and the calculation of natural bond orbitals~\cite{Ohwaki2014}.  This latter capability was used to analyze a molecular dynamics simulation on a liquid electrolyte bulk model, namely propylene carbonate + \chemform{LiBF_4} in a model containing 2176 atoms~\cite{Ohwaki2014}.

\TD{%
\paragraphlb{FHI-aims} %https://aimsclub.fhi-berlin.mpg.de
The Fritz-Haber-Institute \emph{ab initio} molecular simulations package~\cite{Blum2009,fhi-aims_website} uses explicit confining potentials to construct numerical atom-centered orbital basis functions;
around $50$ basis functions per atom are needed to have an accurate solution of less than one meV per atom.
%\textcolor{blue}{[IS THE FOLLOWING A MERGE PROBLEM??]}
%>>>>>>> refs/remotes/origin/master
This scheme can be used naturally to achieve quasi-linear scaling for the grid based operations~\cite{Havu2009} with a demonstrated
$O(N^{1.5})$ overall scaling for a linear system of polyalanine up to $603$~atoms. The authors of FHI-aims have also developed a massively parallel eigensolver, ELPA~\cite{Marek2014}, for large dense matrices based on a two-step procedure (full matrix to a banded one, and banded matrix to
a tridagonal one). 
Traditional DFT and embedded-cluster DFT~\cite{Berger2014} calculations can be done on molecules~\cite{Schubert2015}, but also on periodic systems.
Hybrid functionals~\cite{Schubert2015}, RPA, MP2, and GW methods are also implemented using a resolution of identity~\cite{Ren2012a} based on auxiliary basis functions.
}

\paragraphlb{CONQUEST} %http://www.order-n.org 
\textsc{Conquest}~\cite{Bowler2000,Bowler2002,Bowler2010,conquest_website} uses an approach based on support functions and density matrix minimization. The support functions can be represented either in a systematic B-spline basis, or in a basis of PAOs, according to the user's preference.  There is also a choice of using a linear-scaling approach wherein the density matrix is optimized using LNV or a cubic-scaling approach using diagonalization. Constrained DFT~\cite{Sena2011} and multisite support functions are implemented, wherein the support functions are associated with more than one atom~\cite{Nakata2014a}.  
Scaling tests have been performed on up to 2~million atoms of bulk Si~\cite{Bowler2010} and the approach has also been applied to Ge hut clusters on Si, 
for systems of up to $23,000$ atoms~\cite{doi:10.1143/JPSJ.77.123706}.
Other examples of calculations with \textsc{Conquest} include $3400$ atom simulations of hydrated DNA~\cite{Otsuka2008} and molecular dynamics (MD) simulations of over $30,000$ atoms of crystalline Si~\cite{Arita2014} using the extended Lagrangian Born-Oppenheimer method~\cite{Niklasson2008}.
     
\paragraphlb{BIGDFT}
The \textsc{BigDFT} code~\cite{bigdft_website} emerged as an outcome of an EU project in 2008.
One of the most particular features of this code is the basis set it uses, Daubechies wavelets~\cite{daubechies1992ten}. These functions have the remarkable property of --- at the same time --- being orthonormal, having compact support in both real and reciprocal space and forming a complete basis set. 
Such a basis set offers optimal properties for DFT at large scale.
The code was first designed following a traditional cubic-scaling approach~\cite{Genovese2008}, and later complemented with a linear-scaling algorithm~\cite{mohr-daubechies-2014,mohr-accurate-2015}. Since wavelets form a very accurate basis set, \textsc{BigDFT} is --- in conjunction with elaborate pseudopotentials --- capable of yielding a very high precision\cite{Willand2013} at maintainable computational costs. This is also true for the linear-scaling version, where the support functions are expanded in the wavelet basis and can thus be adapted \emph{in situ}. The main approach used to optimize the density matrix and thereby achieve linear scaling is Fermi Operator Expansion~\cite{Goedecker1999}.

Some features implemented in \textsc{BigDFT} are, among others, time dependent DFT~\cite{Natarajan2012} and constrained DFT, which has been implemented based on a fragment approach~\cite{Ratcliff2015Fragment}, along a similar spirit to the fragment molecular orbital approach described in more detail below. In addition \textsc{BigDFT} incorporates a very efficient Poisson Solver based on interpolating scaling functions\cite{Genovese2006,Genovese2007,Cerioni2012,Fisicaro2015}, which solves the electrostatic problem with a low $\mathcal{O}(N \log N)$ complexity and a small prefactor and can thus also be used for large scale applications. \textsc{BigDFT} was also one of the first DFT codes taking benefit of accelerators used in HPC systems,
such as Graphic Processing Units~\cite{Genovese2009}. 
Some of the code developers are among the authors of the present review, therefore some illustrative examples that will be given in the following sections originate from runs with \textsc{BigDFT}.

\paragraphlb{ERGOSCF} 
\textsc{ErgoSCF}~\cite{Rudberg2011,ergoscf_website} %\textcolor{blue}{[GH: repeating all those names looks strange, I'd suggest a linebreak after the paragraph title or rephrasing. SM: added a line break everywhere]} 
is a quantum chemistry code for large scale HF and DFT calculations, which has a variety of pure hybrid functionals available and is an all-electron approach based on Gaussian basis sets.  It uses a trace-correcting purification method in conjunction with fast multipole methods, hierarchic sparse matrix algebra, and efficient integral screening to achieve linear scaling.  It has been applied to protein calculations, using both explicit and implicit solvents~\cite{Rudberg2012}.

\paragraphlb{FREEON} 
Formerly \textsc{mondoscf}, \textsc{FreeON}
is a suite of linear-scaling experimental chemistry programs~\cite{freeon_website} which performs HF, pure DFT, and hybrid HF/DFT calculations in a Cartesian-Gaussian LCAO basis. All algorithms are $O(N)$ or $O(N \textrm{log} N)$ for non-metallic systems.
Different purification and density matrix minimization approaches have been implemented and compared in the code~\cite{Jordan2005}.

\paragraphlb{QUICKSTEP} 
The quantum mechanical part of the \textsc{CP2K}~\cite{cp2k_website} package, \textsc{Quickstep}~\cite{VandeVondele2005},
uses traditional Gaussian basis sets to expand the orbitals, whereas the electronic density is expressed in plane waves
to perform HF, DFT, hybrid HF/DFT and MP2~\cite{Ben2015} calculations. 
The Kohn-Sham energy and Hamiltonian matrix is calculated using a linear-scaling approach with screening techniques~\cite{CP2K.2005}.

\paragraphlb{FEMTECK} 
The Finite Element Method based Total
Energy Calculation Kit \textsc{femteck} code~\cite{Tsuchida1998,Tsuchida2007} uses an adaptive finite element basis to represent Wannier functions, in conjunction with the augmented orbital minimization method (OMM), which imposes additional constraints on the orbitals to guarantee linear independence in order to overcome the slow convergence and local minima usually associated with the standard OMM method.  \textsc{femteck} has been used for molecular dynamics simulations of liquid ethanol with $1125$ atoms~\cite{Tsuchida2008}, as well as for the study of fast-ionic conductivity of \chemform{Li}~ions %\textcolor{red}{[SM: shouldn't the + be raised, like \ce{Li+}?]}
in the high-temperature hexagonal phase of \chemform{LiBH_4}, in MD simulations of $1200$ atoms~\cite{Ikeshoji2011}.

\paragraphlb{RMGDFT}
The real space multigrid based DFT electronic structure code~\cite{rmgdft_website}
(\textsc{RMGDFT}) uses a multigrid~\cite{Fattebert2000} or a structured non-uniform mesh~\cite{Fattebert2007}.
A linear-scaling method~\cite{Fattebert2004} was also developed. Using maximally localized Wannier functions expressed on a uniform finite difference mesh, Osei-Kuffuor and Fattebert~\cite{Osei-Kuffuor2014} performed a molecular dynamics simulation up to $101,952$ atoms of polymers to demonstrate the scalability of their algorithms.

\paragraphlb{PROFESS}
As the imposition of orthonormality constraints on the KS orbitals is one of the factors which dominates the cubic-scaling of standard DFT, one strategy to achieve linear-scaling is to eliminate the need for the orbitals.  This so-called orbital-free (OF) DFT approach does so by defining a kinetic energy (KE) functional, for which several forms have been proposed, see e.g.\ Refs.~\onlinecite{Wang1992,Garcia2007,Huang2010}. Such an approach has been implemented in \textsc{profess} (Princeton Orbital-Free Electronic Structure Software)~\cite{Ho2008,Hung2010,Chen2015,profess_website}, which offers a choice between several implemented KE functionals using a grid-based approach to represent the density.  The code requires the use of local pseudopotentials, which are provided for certain elements only, including  Mg, Si and Al.  The approach only achieves the same accuracy as KS-DFT for main group elements in metallic states, but recent work developing new KE functionals for semiconductors and transition metals~\cite{Huang2010,Shin2014,Chen2012} allow some properties of semiconductors to also be reproduced well.  Despite this limitation, large defects in crystals (e.g., dislocations, grain boundaries) and large nanostructures (e.g., nanowires, quantum dots) are too computationally costly to treat with most first principles approaches, and so OF-DFT offers an appealing alternative for such systems.  The code has been used to simulate more than 1~million atoms of bulk Al~\cite{Hung2009} and to study melting of Li using molecular dynamics~\cite{Chen2013}.
 
\paragraphlb{Quantum chemistry}
Reduced-scaling electronic structure methods are a domain where the ultimate goal is to have a linear-scaling approach with chemical accuracy. Based on pair natural orbitals, a coupled cluster theory method~\cite{Riplinger2016} has been developed which scales up to $1000$ atoms claiming that chemical accuracy was achieved.
A Quantum Monte Carlo method is also being developed for large chemical systems with some calculations on peptides~\cite{Scemama2013,Scemama2013a} up to $1731$~electrons.

\paragraphlb{Machine learning} 
Finally we wish to mention the various works on neural networks and other machine learning techniques where the goal is to obtain interatomic potentials with the same accuracy as DFT or even quantum chemistry. The first works~\cite{Behler2007,Behler2008,Behler2008a} of J.\ Behler  using a high-dimensional neural network give a way of calculating potential-energy surfaces in order to perform metadynamics.
Another approach is to use the electronic charge density coming from DFT to build interatomic potentials for ionic systems~\cite{Ghasemi2015}. By using GPUs it is possible to speed up the neural network performance by two orders of magnitude, which permits a large computing capacity within a single workstation.
%\textcolor{blue}{
Rupp et al.\ considerably improved the predictive precision and transferability of spectroscopically relevant observables and atomic forces for molecules using kernel ridge regression~\cite{Rupp2015}. 
Once such a system is trained, the cost for new calculations is orders of magnitude smaller than for corresponding DFT calculations.
%}

\subsection{Towards coarse-graining modelling of large systems: FMO and DFTB approaches}
\paragraphlb{Fragment Molecular Orbitals} One important method for proteins and other biological molecules, which could be considered an extension of the divide-and-conquer approach, is the fragment molecular orbital (FMO) approach~\cite{Kitaura1,Fedorov2007}.  In this approach, the molecule is divided into fragments---whose definition is based on chemical intuition---, which are each assigned a number of electrons.  The size of each fragment might therefore vary depending on the system in question, for example 2D pi-conjugated systems would need large fragments for accuracy.  The molecular orbitals (MOs) are then calculated for each fragment, under the constraint that they remain localized within the fragment.
The distinguishing feature of FMO compared with divide and conquer is that the MOs for the fragments are calculated in the Coulomb field coming from the rest of the system (i.e.\ the environment), so that long range electrostatics are included. The fragment MOs must be updated iteratively to ensure self-consistency of this environment electrostatic potential. 
Different levels of approximation can be used: the most basic is FMO1, which only explicitly calculates MOs of single fragments (referred to as monomers) and constructs the total energy from these results.  The next, and most common level, is FMO2, which also incorporates explicit dimer calculations (i.e.\ between pairs of fragments) into the total energy.
There is also FMO3~\cite{Fedorov2004}, which also adds trimers, and even FMO4, which incorporates also 4-body terms~\cite{Nakano2012}.  The accuracy of the approximation, but also the cost increases with the addition of higher order terms.
FMO2 is often sufficiently accurate for many applications, but there are some cases where higher order interactions are required, for example, at least three-body terms were found to be necessary for MD of water~\cite{Pruitt2016}; geometry optimizations of open shell systems may also require higher order terms, or, where possible, larger fragments in order to ensure good convergence~\cite{Pruitt2012}.
FMO has been implemented in \textsc{GAMESS} (General Atomic and Molecular Electronic Structure System)
~\cite{Schmidt1993,Gordon2005,gamess_website}, with an implementation which is designed to exploit massively parallel machines; \textsc{ABINIT-MP}~\cite{Nakano2006,abinitmp_website}, which has also been designed for massively parallel calculations~\cite{Mochizuki2008}; and a version of \textsc{NWChem}~\cite{Sekino2003}.

FMO belongs to a wider class of fragment-based methods, such as the molecular tailoring approach~\cite{Ganesh2006}.  FMO, molecular tailoring and other related approaches have been reviewed in detail elsewhere, along with a number of example applications~\cite{Fedorov2007,Gordon2009,Gordon2012,Pruitt2014}.  Here we highlight a few examples for large systems.  Aside from DFT, FMO may also be used for HF and MP2 calculations; for example, the GAMESS implementation has been benchmarked for water clusters containing around 12,000 atoms at the MP2 level of theory~\cite{Fletcher2012}.  FMO has also been used for geometry optimizations of large systems, for example the prostanglandin synthase in complex with ibuprofen, containing around 20,000 atoms, was optimized using B3LYP and restricted HF (RHF) for different domains of the system~\cite{Fedorov2011}.
Another example application is the study of the influenza virus hemagglutinin, where QM calculations of up to 24,000 atoms~\cite{Sawada2010,Sawada2010a} have been performed using FMO-MP2 in combination with the polarizable continuum model (PCM)~\cite{Fedorov2006,Barone1997}. More than 20,000 atoms were also included in a RHF simulation of the photosynthetic reaction center of rhodopseudomonas viridis, which required around 1400 fragments~\cite{Ikegami2005}.  While less common, FMO can also be applied to solids, surfaces and nanomaterials.  For example a new fragmentation scheme for fractioned bonds was developed and applied to the adsorption of toluene and phenol on zeolite~\cite{Fedorov2008}; Si nanowires have also been studied using FMO~\cite{Fedorov2009}.  FMO may also be used for excited calculations, for example in combination with TDDFT, which has been tested for solid state quinacridone~\cite{Fukunaga2008}.

\paragraphlb{DFTB}
The Density Functional Tight-Binding approach was first notably applied in carbon-based systems. The idea was, at the first-order, to use a frozen density from atoms. DFTB, at the second order, has been extended in order to include a Self-Consistent Charge (SCC) correction~\cite{Elstner-Frauenheim.1998}, accounting for valence electron density redistribution due to the interatomic interactions.
A third-order DFTB3~\cite{Nishimoto2015a} was also developed which has introduced an additional term with coupling between charges.
Parameters for the whole periodic table are available~\cite{Wahiduzzaman2013}. A confinement potential was used to tighten the Kohn-Sham orbitals.
The solution conformations of biologically mono- and di-$\alpha$-D-arabinofuranosides were investigated~\cite{Islam2012} by means of molecular dynamics using dispersion-corrected self-consistent DFTB and compared to the results from the AMBER \SM{ff99SB} force field~\cite{hornak-2006-comparison} \SM{with the GLYCAM (version 04f) parameter set for carbohydrates~\cite{woods-1995-molecular,case-2005-the-amber}} \delete{\cite{Salomon-Ferrer2013}} as well as to NMR experiments. 
%\textcolor{red}{LR: add amber citation? SM: I added a reference, but there are several ones... add still more... TD, one is enough!}
There are also some extensive tests on hydroxide water clusters and aqueous hydroxide solutions~\cite{Choi2013}.

\paragraphlb{FMO-DFTB} Recently, the fragment molecular orbital approach has been combined with DFTB~\cite{Nishimoto2014}, with the aim being to reduce the cost of DFTB to a few seconds, in order to perform MD simulations.  The accuracy of FMO-DFTB is very close to that of DFTB, while excellent speedups have also been achieved: for an MD simulation of 768 atoms of water, the speedup compared to DFTB was shown to be more than 100~\cite{Nishimoto2015b}.
It has been used to optimize an 11,000 atom nanoflake of cellulose $I\beta$~\cite{Nishimoto2015a}, as well as for MD simulations of liquid hydrogen halides containing 2000 atoms~\cite{Nishimoto2015b}, for which the speedup was an order of magnitude greater than the above example.  FMO-DFTB could therefore be a very promising approach of QM-MD of large systems.

\subsection{HPC concepts/performance}

The effort for the development of the above mentioned computer codes has also contributed to another improvement in the community: the ability to exploit high performance computing (HPC) resources. This has become a very important aspect with the advent of petaflop supercomputers. New science can be done on these machines only if the code developers are able to profit from such large scale supercomputers.
However, the development of accelerated methods for large systems is not meant to replace the exploitation of powerful HPC platforms, rather the two go hand in hand: in order to execute calculations for systems as large as the range for which they are designed, LS-DFT codes require large computing resources; and parallel compute clusters are most efficiently exploited if they are used to treat large problem sizes, rather than to compensate for the cubic scaling of standard DFT codes.  

In the ideal case, for a method which exhibits both perfect linear scaling with respect to the number of atoms and ideal parallel scaling with respect to the number of computing cores, the time taken for a single point calculation should remain constant if the ratio of atoms to cores remains constant, that is, a so-called weak scaling curve would be flat.  This allows for the definition of the concept of \emph{CPU minutes per atom} and, correspondingly, \emph{memory per atom}.  Since both the time and memory requirements for a small system running on a few cores would be approximately the same as a large system running on many cores, this value for a particular code is a function of the system, which depends on the dimensionality of the material, the atomic species and various user-defined quantities, such as the grid spacing and the localization radii beyond which localized basis functions are truncated.  

In other terms, we might say that the values of per-atom computing resources for a given code are functionals of the input parameters of the code and of the computing architecture employed. However, even though they cannot be predicted beforehand and have to be evaluated, it is interesting to compare values of CPU minutes and memory per atom between different systems. This is especially useful because such a viewpoint provides a quantitative method for estimating the cost of a large simulation based on a small representative calculation, i.e.\ a smaller but equivalent simulation domain running on the same computing architecture.  For example, when one has a fixed number of cores available, one could estimate the total run time for a large calculation using the CPU minutes per atom value obtained from the small calculation.  Alternatively, in the case where one has many cores but limited memory availability, one could use the memory per atom value from the small calculation to determine the minimum number of cores needed to fit the large problem in memory.  Although in practice one might not achieve perfect parallel scaling, the validity of these quantities has been demonstrated in the context of the \textsc{BigDFT} code, where it has been tested for DNA fragments and water droplets~\cite{mohr-accurate-2015}.  A similar concept was also demonstrated in the context of coupled cluster theory~\cite{Riplinger2016}.

It is however fundamental that all these performance achievements come together with the \emph{robustness} of the approach, or more precisely its implementation. It is easy to imagine that QM methods become technically very complicated at large scale, with a multiplication of the input variables and troubleshooting techniques which are typical of the algorithm employed. Code developers have to provide robust and reliable algorithms for non-specialists (failsafe mode), even at the cost of lower performance. 

\section{Large-scale QM applications}\label{secQMMM}
As already mentioned, first principle calculations are \emph{a priori} the most accurate approach to any atomistic simulation. 
Unfortunately an exact analytical or computational solution to the fundamental quantum mechanical equations is only possible for a few, rare cases at present; for all other systems one either has to introduce approximations, solve the equations numerically, or both. Due to these approximations there might thus even be situations where an empirical approach, which is tuned for one particular property, might yield more precise results than a first principles calculation. One particular, but important example is water, where traditional \emph{ab initio} approaches like DFT do not come as close to the experimental values (see, for instance, Ref.~\onlinecite{Zen2015} and references therein) as empirical force fields~\cite{Vega2011,Kiss2012}. % (see, for instance, Ref.~\onlinecite{Zen2015} and references therein \textcolor{red}{[SM: Shall I give the refs therein (would have to read them first...)?]}.
On the other hand such empirical approaches will only work for systems which are very similar to the ones which were used for the tuning and will in general fail for systems which are distinct, making the simulation of unknown materials tricky. In addition, traditional force fields do not in general allow for bond breaking and forming, which is however abundant in chemical reactions. This is in strong contrast to \emph{ab initio} approaches, which are less biased and are thus expected to yield the correct tendencies over a much larger range of systems.

Moreover there are situations where a first principles description is not only desirable, but essential. Obviously this is the case when quantities are needed which are not accessible with classical force field approaches.
For instance, a major shortcoming of classical approaches is that they cannot provide direct insights into electronic charge rearrangements. This is however necessary if one wants to analyze charge transfers, which play an important role for instance in biology.
Since such electronic charge transfers can occur over a large distance and over a long time frame, such simulations would quickly go beyond the scope of pure \emph{ab initio} calculations. A possible solution is to let the system evolve according to a classical approach, which is orders of magnitudes faster, and only analyze certain snapshots or averages on a QM level. An example is the work of Livshitz \emph{et al.}~\cite{Livshits2014}, which investigates the charge transfer properties of DNA molecules adsorbed onto a mica surface, or the work of Lech \emph{et al.}~\cite{Lech2013} which investigates the electron-hole transfer in various stacking geometries of nuclear acids.
With respect to DNA, \emph{ab initio} calculations have also been used to investigate the molecular interactions of nucleic acid bases~\cite{Sponer2002} and to study the impact of ion polarization~\cite{Gkionis2014}.
A nice overview over the various approaches for DNA can be found in Ref.~\onlinecite{Dans2016}.

The volume of an atom is also an example of such an intrinsic QM quantity, which is most straightforwardly defined using its charge distribution, but can not be accessed directly using force fields; consequently other models must be adopted~\cite{Gaines2015}. Another case where first principles calculations are needed is the determination of photophysical and spectroscopic properties. These calculations do not only require the determination of the ground state, but also also of excited states, which is not possible with classical approaches based on empirical force fields. This application is also very demanding from the point of view of the QM method, since popular approaches like HF or DFT are ground state theories and therefore usually give rather poor results for excited states. A popular solution to this problem is to use TDDFT for the calculation of the excitations~\cite{Natarajan2012}.
  
Highly accurate QM methods are also required in the determination of small energy differences, for instance, the calculation of activation energy barriers in chemical reactions. The problem is that, in particular for biological systems, the reaction is often catalyzed by an environment which can be much larger than the actual active site, making a fully \emph{ab initio} treatment impossible. However, if there is no charge transfer between the active site and its environment, it is possible to use so-called QM/MM schemes, where only the active site is treated on a highly accurate \emph{ab initio} level and the environment is handled using force fields. As an example, such an approach was used, among others, to investigate catalysts for the Kemp eliminase~\cite{Kiss2010}, and QM calculations in general are an important ingredient for the computational design of enzymes~\cite{Kiss2013}. 

Obviously such a QM/MM strategy raises the question of how the coupling between QM and MM regions can be done; \LG{see Ref.~\onlinecite{SubsystemDFT} for an interesting review on the subject}. Typically one distinguishes between three different setups, namely mechanical embedding, electronic embedding and polarized embedding~\cite{bakowies1996ahierarchy,Lin2007,Senn2009}. In the first case, the interaction between the QM and MM region is treated in the same way as the interaction within the MM region itself; in the second case, the MM environment is incorporated into the QM Hamiltonian, thus leading to a polarization of the electronic charge density; and in the third case, the MM environment is also polarized by the QM charge distribution. The second method is the most popular one, and the coupling between QM and MM region can for instance be done using a multipole representation which is fitted to the exact electrostatic potential~\cite{Ferre2002}.

Unfortunately electronic embedding is known to exhibit the shortcoming of ``overpolarization'' at the boundary between the QM and MM region~\cite{Senn2009,Neugebauer2009}\SM{, in particular} if covalent bonds are cut.
\SM{This overpolarization problem is due to the fact that for the electrons of the MM atoms the Pauli repulsion is not accounted for, resulting in an incorrect description of the short range interaction at the QM/MM interface. In particular, positive atoms on the MM side might act as traps for QM electrons, leading to an excessive polarization.
However there exist several approaches to address this issue, for instance the use of a delocalized charge distribution for the MM atoms~\cite{eichinger-1999-a-hybrid,das-2002-optimization,biswas-2005-a-regularized}, and indeed it can be shown that a careful implementation allows the correct description of polarization effects within a QM/MM approach~\cite{senthilkumar-2008-analysis}.}
Another, more straightforward solution to this problem is to increase the size of the QM region, keeping the problematic boundary farther away from the active site. Since in this way the QM region easily contains hundreds or thousands of atoms, the use of a linear-scaling algorithm for the QM part is indispensable.
In a \textsc{onetep} application by Zuehlsdorff \emph{et al.}~\cite{Zuehlsdorff2016} they even found that an explicit inclusion of the solvent into the QM region was required to get a reliable description.

In another recent work employing \textsc{onetep}, Lever \emph{et al.}~\cite{Lever2014} used this LS-DFT code to investigate the transition states in enzyme-catalyzed reactions.
\SM{The same reaction has already been investigated earlier with a QM/MM approach employing for the QM part highly accurate quantum chemistry methods such as MP2, LMP2, and LCCSD(T0)~\cite{claeyssens-2006-high-accuracy}. Even though the development of reduced scaling algorithms~\cite{schuetz-1999-low-order-I,hetzer-2000-low-order-II,schutz-2000-low-order-III,schuetz-2001-low-order-IV,schuetz-2002-low-order-V,schuetz-2000-local,werner-2003-fast} made their use somewhat affordable also for larger systems, the cost of these methods is still very high. On the other hand they have the advantage that they yield very accurate estimates for activation barriers, in contrast to DFT, which in general tends to underestimate these values.
Indeed another study by Ml\'ynsk\'y \emph{et al.}\ demonstrated the need for using appropriate methods for the QM treatment within a QM/MM approach. Whereas all methods (quantum chemistry, DFT and semi-empirical) gave similar reaction barriers, the reaction pathways were considerably different for the semi-empirical calculations~\cite{mlynsky-2014-comparison}.
Recent developments also allow the embedding of a small region which is treated by quantum chemistry methods within a larger region which is treated by DFT, and both regions can then also be used within a QM/MM approach~\cite{manby-2012-a-simple,bennie-2016-a-projector}.}

\SM{Instead of falling back to expensive quantum chemistry methods it is also possible to improve the accuracy of DFT calculations in a cheap way by including dispersion corrections. A study by Lonsdale \emph{et al.}\ found that this considerably improved the values of the calculated energy barriers~\cite{lonsdale-2012-effects}. Generally speaking, it is recommendable to include such dispersion corrections in any large scale QM calculation, as they add only a small overhead, but may improve the physical description considerably.}

The QM/MM philosophy is also useful to calculate, for a given subsystem, quantities which are intrinsically only possible with an \emph{ab initio} approach, but influenced by a surrounding which does not require a strict first principles treatment.
For instance, excitation energies and the absorption spectrum of DNA were calculated by Spata et al.~\cite{Spata2014} using an electrostatic embedding QM/MM approach and by Gattuso et al.~\cite{Gattuso2015} using a QM/MM approach based on the Local Self Consistent Field (LSCF) method~\cite{Monari2013}.
\LG{Also heavy atoms like actinides can be considered in this scheme \cite{UranylEmbedded}.}

\SM{Additionally, as QM/MM calculations try to couple various levels of theory}\LG{, see e.g.\ Ref.~\onlinecite{MichelFerre}}\SM{, it might be necessary in some cases to manually adjust some force field parameters, as for instance done by Pentik\"ainen \emph{et al.}\ for the QM/MM simulation nucleic acid bases~\cite{pentikainen-2009-lennard-jones}, and to carefully check the compatibility of the chosen methods~\cite{shaw-2010-compatibility}.
For some applications it might even be the case that a ``traditional'' QM/MM approach (i.e.\ involving two levels of description) is not sufficient in order to cover the entire length scale. Thus one might have to use additional levels of coarse graining and abstraction, together with a coupling between them, as has for instance been done by Lonsdale \emph{et al.}~\cite{lonsdale-2014-a-multiscale}}

\SM{To summarize, QM/MM approaches seem to give, at least qualitatively, very useful results, and the main source of error is rather due to a lack of physical correctness in the QM model than in the QM/MM partitioning.}

\begin{figure}
 \includegraphics[width=0.5\textwidth]{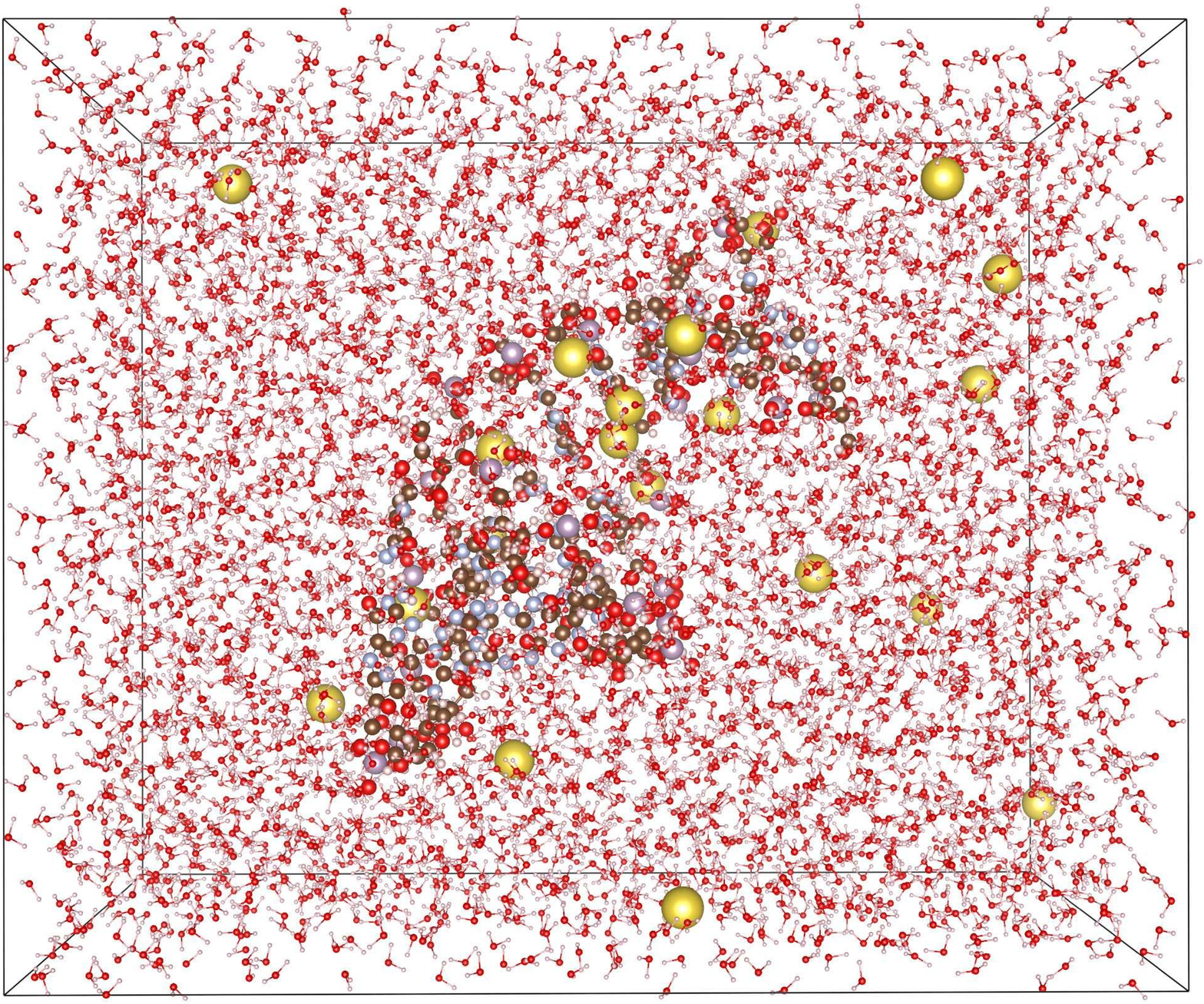}
 \caption{Visualization~\cite{Momma2011} of a DNA fragment containing 11 base pairs,
 surrounded by a solvent of water and Na ions (giving in total 15,613 atoms), with periodic boundary conditions. }
 \label{fig:DNA_water-Na}
\end{figure}

\begin{figure}
 \includegraphics[width=0.5\textwidth]{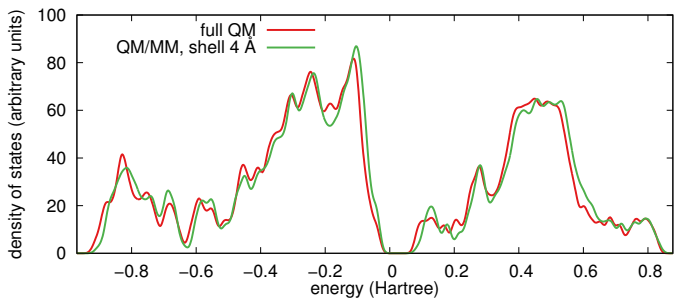}
 \caption{Partial density of states for the DNA within the system depicted in Fig.~\ref{fig:DNA_water-Na}. The red curve was generated treating the entire system on a QM level, whereas the green curve only treated the DNA plus a shell of \unit[4]{\r{A}} on a QM level, with the remaining solvent atoms replaced by a multipole expansion. In order to allow for a better comparison, the QM/MM curve was shifted such that its HOMO energy coincides with the one of the full QM approach.}
 \label{fig:DNA-water_QMMM_pdos}
\end{figure}

The calculation of the partial density of states is another example intrinsically requiring a QM treatment, which we will demonstrate for the system depicted in Fig.~\ref{fig:DNA_water-Na}, showing a small fragment of DNA in a water-Na solution consisting in total of $15,613$~atoms.
The determination of the electronic structure is only possible using a QM method, but the influence of the environment on the DNA can also be modelled with a less expensive classical approach. In Fig.~\ref{fig:DNA-water_QMMM_pdos}, we compare the outcome of a full QM calculation with a static QM/MM approach, where all the solvent except for a small shell around the DNA has been replaced by a multipole expansion up to quadrupoles, leaving in total only $1877$~atoms in the QM region. 
Both calculations were done with \textsc{BigDFT}, and the multipoles were calculated as a post-processing of the full QM calculation.
As can be seen from the plot, the two curves are virtually identical, but the QM/MM approach had to treat about 8 times fewer atoms on a QM level and was thus computationally considerably cheaper.

Another field of application for QM methods is the parametrization of force fields. Many of the widely used force fields are fitted to reproduce experimental data of a certain test set of structures. However the outcome of this fitting procedure is not necessarily transferable to other compounds~\cite{Mackerell2004}.  A more severe problem is the lack of applicability to different physicochemical conditions, such as pressure or temperature. This approach might thus lead to bad results when these force fields are applied to systems or conditions which are considerably different than the ones used for the parametrization. A possible solution is to parametrize a force field using results from \emph{ab initio} calculations, which widens the range of possible applications.\delete{For instance, atomic charges derived from the restrained electrostatic potential (RESP) approach~\cite{Bayly1993} have been used to parametrize \SM{certain versions of} the AMBER force field, with very promising results~\cite{Wang2000}\SM{\cite{cornell-1995-a-second}}.}
\SM{For instance, certain versions of the AMBER force field have been parametrized using atomic charges derived from \emph{ab-initio} calculations, as for instance those described by  Weiner \emph{et al.}~\cite{weiner-1984-a-new}, Cornell \emph{et al.}~\cite{cornell-1995-a-second} or Wang \emph{et al.}~\cite{Wang2000}; for the last two, charges derived from the restrained electrostatic potential (RESP) approach~\cite{Bayly1993} have been used.}
\emph{Ab initio} results were also included---among experimental results---into the parametrization of the CHARMM22 force field~\cite{MacKerell1998}. There have also been attempts to develop force fields which determine the optimal set of parameters in an automatic way, using \emph{ab initio} results as target data~\cite{Huang2013}.
A logical continuation of this line uses statistical learning and big data analytics as envisioned in the European project NOMAD~\cite{nomad_website} which has the goal of using these techniques on top of a large computational material database.

\begin{figure}
\includegraphics[width=0.5\textwidth]{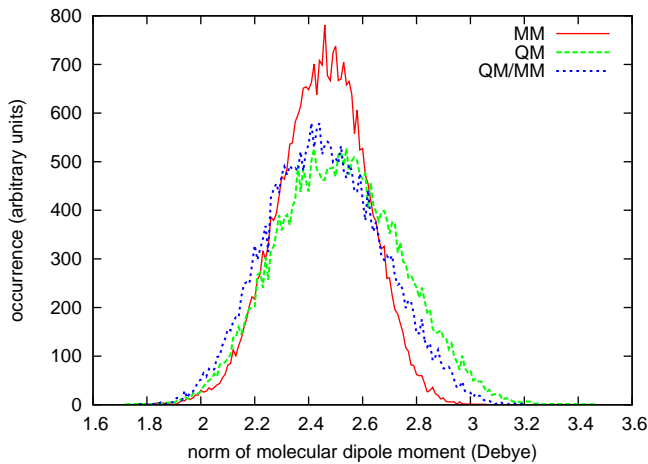}
 \caption{Dispersion of the molecular dipole moment of water molecules within a droplet of 1800 atoms, with statistics taken over 50 snapshots of an MD simulation. The dipole is calculated based on the atomic monopoles and dipoles, and these were obtained from a) a classical simulation using \textsc{POLARIS(MD)}, b) a DFT simulation using \textsc{BigDFT}, and c) a combined QM/MM approach.}
% \textcolor{blue}{[GH: Obviously a histogram, still y-axis missing]} \textcolor{green}{[SM: I added "occurrence"... any better suggestions?]}}
 \label{fig:water_comparison_dipoles}
\end{figure}

Finally we also highlight the advantages of \textit{large scale} QM simulations. 
Sometimes one is interested in atomistic characteristics averaged over a large number of samples, in this way generating the macroscopic behavior. An example is the dipole moment of liquid water, which is a macroscopic observable with a microscopic origin.
In order to calculate it accurately it is not sufficient to simply compute the dipole moment of one water molecule in vacuum. Instead one has to take into account the polarization effects generated by the other surrounding water molecules.
Due to thermal fluctuations each molecule will however yield a different value, and the macroscopic observable result (keeping in mind that this can only be determined indirectly and is thus itself subject to fluctuations) can therefore only be obtained by averaging over all molecules, thereby requiring a truly large scale first principles simulation. The outcome of such a simulation, carried out using the MM code \textsc{POLARIS(MD)}~\cite{Real2013} and the QM code \textsc{BigDFT} is shown in Fig.~\ref{fig:water_comparison_dipoles}. Here we plot the dispersion of the molecular dipole moments, calculated based on atomic monopoles (i.e.\ atomic charges and dipoles) of a water droplet consisting of 600 molecules at ambient conditions and taking 50 snapshots of an MD simulation. As can be seen, there is a wide dispersion of the molecular dipole moments, which however yield a mean value in line with other theoretical and experimental studies~\cite{gubskaya2002for}.% the experimentally observed value of about \unit[2.5??]{Debye}.

%\fbox{
%\begin{minipage}[t]{0.5\textwidth}
\section{Multiscale linked together: an example} \label{extmossec}
The above presented studies, linking together different models and length scales, are of course only a small set of representatives of the ongoing works in the literature.
The need for a connection of models in the common scale regimes is not only
related to the  Quantum Mechanical/semiclassical regime.
Such multi-method schemes can be applied also to larger length scales, up to sizes 
of interest for actual industrial applications.
There is therefore a direct implication of large-scale QM methods on present-day \emph{technological} challenges.

\begin{figure}
\includegraphics[width=0.5\textwidth]{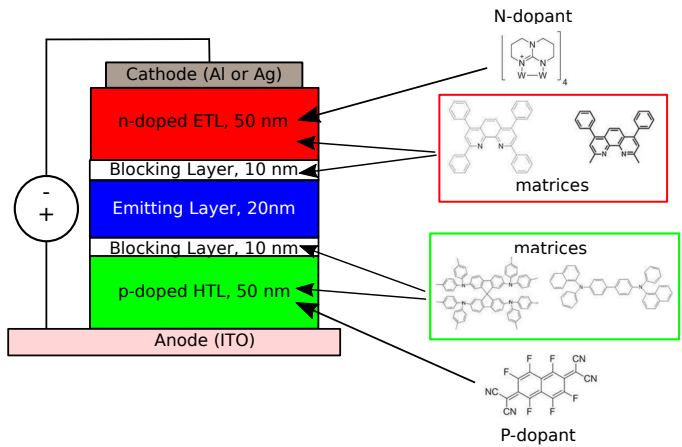}
\caption{\TD{Organic light-emitting diodes (OLED) device configuration illustrating the target goal of the EXTMOS project: simulating the full device from the molecular composition of the different layers.}} 
%\textcolor{blue}{[GH: Very small font size for the chemical composites]}
%\textcolor{magenta}{[TD: Change the fonts.]}
%}
\label{fig:extmos}
\end{figure}
\begin{figure}
\includegraphics[width=0.5\textwidth]{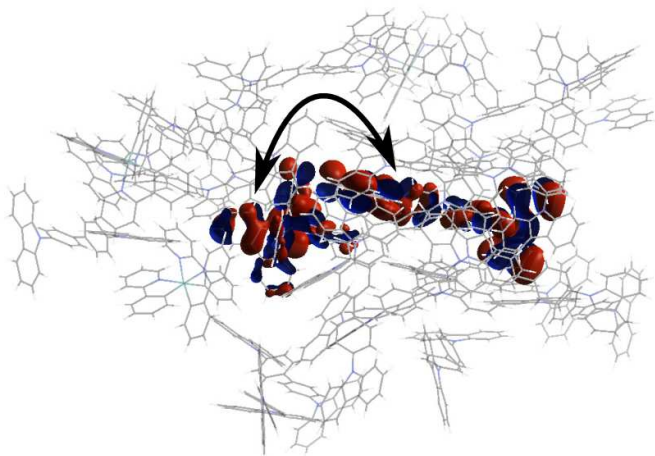}
\caption{Plot showing the HOMOs of two neighboring molecules calculated using a fragment approach.  Their nearest neighbors extracted from a large disordered host-guest morphology are also depicted.  Using this setup, one can calculate transfer integrals which take into account the environment~\cite{Ratcliff2015}.}
\label{fig:oled}
\end{figure}

As an illustrative example we present the 
European project H2020 EXTMOS. The objective of this project is to build a model simulating
organic light emitting diodes (OLEDs) in order to calculate their efficiency, where the only inputs are the organic molecule components. 
In the OLED realization process, acceptor and donor molecules, together with some dopant molecules,
are mixed in a thin film; see Fig.~\ref{fig:extmos}.
It can be easily imagined that the investigation of such a process requires a multi-scale 
approach with a coupling between different description levels.
We will briefly describe them here.
\paragraph{Phase organisations}
Calculating the morphology of the organic film is the first step, which is done by means of molecular mechanics or molecular dynamics at room temperature
using an appropriate polarizable force field~\cite{Baker2015} (PFF). 
These PFFs are fitted with a charge analysis coming from DFT in order to reproduce the electrostatic potential.
This step is crucial especially when considering different dopant molecules in the process.
Systems of a few hundred atoms are simulated in QM, calculating the atomic forces and electrostatic potential~\cite{Bayly1993}
which are compared with those coming from PFFs.
As soon as the PFF is fitted, morphologies of the organic film can be calculated and correct statistics of many thousands of 
atoms with their atomic positions can be easily generated even at different temperatures. %\textcolor{red}{[SM: isn't this a contradiction to what Michel said in his mail? LG: I think it is not as Michel was rather referring to different phases whereas here the range in temperature is not likely to modify the phase of OLED constituents (a good display should not freeze or melt ;) ) \textcolor{green}{SM: good point, I added a sentence to explain this (or do you think it is obvious?) LG: I think it is very good, it helps in further clarifying. This message will auto-destroy in few hours...:) }]}
Since the device will only operate within a limited temperature range---in particular only within one phase---the PFF parameters should be transferable without notable loss of accuracy.
\paragraph{Determination of the electronic properties}
As soon as atomic configurations are determined, quantities of interest for the electronic properties of the molecules need to be extracted.
Other QM methods coming from many-body perturbation theory (MBPT) such as 
GW~\cite{Hedin1965} and Bethe-Salpeter methods~\cite{Onida2002} can be used to calculate 
the intrinsic properties of the organic molecules~\cite{Blase2011, Duchemin2012}.
The challenge is to use such methods within an environment, modelled by adequate electrostatic degrees of freedom to describe the morphology of the organic film~\cite{DAvino2014}. 
\paragraph{Hopping integrals}
The previous step permits the calculation of the charge transfer of a few organic molecules in a given embedding environment.
%\textcolor{blue}{[GH: does it allow the calculation only for some molecules or is the charge transfer happening only at some molecules???]}
Since configurational statistics are important to represent correctly an organic film, constrained DFT~\cite{Wu2006} is well suited to understanding the influence of the environmental degrees of freedom~\cite{Schober2015} as well as to impose the correct charge transfer and to calculate the statistic of hopping and site integrals~\cite{Ratcliff2015} over an ensemble of molecules from the morphology (see Fig.\ref{fig:oled} for an illustration).
Here again, another important quantity is the \emph{dispersion} of the results provided by the morphologies. The QM fragment approach
%\textcolor{blue}{[GH: not better "fragment approach for QM calculations"?]}
is well suited to calculate a set of hundreds of molecules in different orientations and environments. 
\paragraph{Towards Device Simulation}
The hopping and site integral parameters are finally used to calculate the efficiency~\cite{Cornil2013} of the organic film
%\textcolor{blue}{[GH: I'd leave out "given by its composition" since it makes this sentence bulky and is already mentioned in the beginning]} 
using a Kinetic Monte Carlo method to predict charge and exciton transport processes through a random walk simulation.
Transport parameters and device characteristics are deduced from the trajectories.
Finally these parameters are included in a drift diffusion simulation in order to simulate larger region sizes and determine a circuit model.

In this example, the role of QM is important to determine correctly the film morphology and also the electronic properties. Nevertheless, QM needs to be used in collaboration with complementary methods: force fields for tractable molecular dynamics and kinetic Monte Carlo methods to deal with larger systems.

%\end{minipage}
%}

%\section{Notable messages} \textcolor{red}{[SM: rather call this "Conclusion"? And I think we should maybe still add some "glue text" to link the various aspects together.]}
\section{Conclusion and outlook}
Advanced atomistic simulation techniques of many different flavors have found widespread applicability during the past years.
Out of this plethora, we have seen the features of some QM codes that are now able to deal with systems with many thousands of atoms.
Most of these techniques were invented more than a decade ago, however the approach to large-scale QM calculations is changing in the present day. 
We might even say that we are entering a ``second era'' of DFT and, more generally, of QM methods in computational science.

On the one hand, the large research effort within the Quantum Chemistry and materials science communities is still ongoing with a focus on small scale systems, trying to achieve very high accuracy (e.g.\ novel exchange-correlation functionals, MBPT methods) and to improve the precision and reliability of the various codes and approaches~\cite{Deltatest}.
However, as this ongoing work concentrates on small scale systems, the QM methods which are nowadays able to arrive at large scales rely on slightly
more mature approaches and are thus forcibly less accurate than state-of-the-art QM methods. 
In other words, the fact that we have nowadays the ability to efficiently treat big systems does not mean that all problems at lower scale are solved.

On the other hand, we have seen a considerable effort of the community to enlarge the accessible length scales of QM simulations. These developments did not aim at developing new approaches to solve the fundamental QM equations, but rather tried to translate existing concepts into new domains. We have seen that this transition was driven by various aspects.

The first important point is related to the reliability of a calculation. One might raise the question whether a QM treatment is still appropriate above ``traditional'' length scales: as already stated, a calculation which is more complex is not necessarily more accurate. But a QM approach is definitely less biased, leading therefore to considerably less arbitrariness. This is in strong contrast to established approaches such as force fields, where the output of a calculation depends strongly on the input of the calculation, for instance the chosen parametrization.
When possible, it is helpful and important to use QM approaches also for large systems, in order to get unbiased insights into the effects of \emph{realistic} experimental conditions on the values of interest, thereby yielding a deeper understanding of fundamental descriptions and trends.
It is therefore important to have the possibility to extend already established QM models to large sizes, in order to have an idea of the effects of such realistic conditions. This leads to a \emph{statistical} approach to large-scale calculations.

These considerations come at hand with the obvious observation that we \emph{have} to abandon the QM treatment above a given length scale where a quantum description will be unnecessary. We used on purpose  ``unnecessary'' instead of ``impossible'' or ``not affordable''; at large scale, a QM calculation is justified \emph{only} if there is the \emph{need} to perform it. There will be no point in obtaining, with a QM treatment, results that could have been obtained with a more compact description like Force Fields or Coarse Grained Models, unless these need to be validated first.
This means that we have to provide strategies to couple the QM description with the modelling methods above this maximum length scale. In other terms, we must be able to provide, eventually, a \emph{reduction} of the complexity of the description, implying that a good QM method at large scale has to provide different levels of theory and precision that can be linked to mesoscopic scales (e.g.\ atomic charges, Hamiltonian matrix elements, basis set multipoles, second principles\cite{Wojdel2015}). 
We have presented in Sec.~\ref{extmossec} one example where such a multi-method approach, completed with a modern QM treatment for the electronic excitations, can lead to results with potential technological implications.

This is also important in those cases where the QM level of theory alone is not able to correctly describe the properties of the system and must be complemented with other approaches.
Thus, the large-scale QM methods described above are important to ``bridge'' the length scale gap with non-QM methods; only if we can perform QM and post-QM approaches for systems with the same size, are we able to see if the trends---if not the actual quantities---are similar, in this way validating the \LG{respective levels of theory}.% post-QM approaches.

A fundamental aspect for this task is the 
\emph{systematicity}
of the investigation. The ability to refine coarse-grained results at a QM level would help at least to \emph{identify} if a refinement of the description might provide different trends. With respect to this task, the diversity of available QM approaches is thus essential.

The approach to large-scale QM calculations is not a mere question of a ``good software''; rather, it represents an opportunity to work in connection with different sensibilities.
This will help in establishing a cross-disciplinary community, working at large scales and connecting together researchers with different sensibilities working with different computational methods and know-how.
This point will be beneficial in both directions. 
Specialists of QM methods will learn to deal with the typical problems related to simulations at the million atom scale, taking advantage of the large experience acquired through the well established classical approaches over the past decades.
For people with a background in classical approaches, tight collaborations with the electronic structure community will offer access to quantities and descriptions that are out of reach without
the sensibility and experience of researchers working in QM methods. %such collaborations are not only desirable, but essential. 

Due to all these reasons the field of large scale QM calculations might attract much attention during the forthcoming years. The topic presents \LG{big} challenges, but offers even greater opportunities.

\section{Acknowledgments}
We would like to thank Modesto Orozco and Hansel G\'omez for fruitful discussions and F\'atima Lucas for providing various test systems and helping with some visualizations.
This work was supported by the EXTMOS project, grant agreement number 646176, and the Energy oriented Centre of Excellence (EoCoE), grant agreement number 676629, funded both within the Horizon2020 framework of the European Union.
This research used resources of the Argonne Leadership Computing Facility at 
Argonne National Laboratory, which is supported by the Office of Science of the 
U.S.\ Department of Energy under contract DE-AC02-06CH11357.

\bibliography{My_Collection}

\onecolumngrid

\end{document}